\documentclass[letterpaper, 10 pt, conference]{ieeeconf}
\usepackage{caption}
\IEEEoverridecommandlockouts
\usepackage{cite}
\usepackage{amsmath,amssymb,amsfonts}
\usepackage{subcaption}
\usepackage{algorithmic}
\usepackage{glossaries}
\usepackage{graphicx}
\usepackage{textcomp}
\usepackage{xcolor}
\def\BibTeX{{\rm B\kern-.05em{\sc i\kern-.025em b}\kern-.08em
    T\kern-.1667em\lower.7ex\hbox{E}\kern-.125emX}}

\usepackage{tikz,xcolor,hyperref}
\definecolor{lime}{HTML}{A6CE39}
\DeclareRobustCommand{\orcidicon}{
	\begin{tikzpicture}
	\draw[lime, fill=lime] (0,0) 
	circle [radius=0.16] 
	node[white] {{\fontfamily{qag}\selectfont \tiny ID}};
	\draw[white, fill=white] (-0.0625,0.095) 
	circle [radius=0.007];
	\end{tikzpicture}
	\hspace{-2mm}
}
\foreach \x in {A, ..., Z}{\expandafter\xdef\csname orcid\x\endcsname{\noexpand\href{https://orcid.org/\csname orcidauthor\x\endcsname}
			{\noexpand\orcidicon}}
}

\title{\LARGE \bf V2P Collision Warnings for Distracted Pedestrians: A Comparative Study with Traditional Auditory Alerts}

\author{Novel Certad\orcidN{} \emph{Graduate Student Member, IEEE}, Enrico Del Re\orcidE{} \emph{Student Member, IEEE}, \\
Joshua Varughese\orcidJ{} \emph{Member, IEEE}, and Cristina Olaverri-Monreal\orcidC{} \emph{Senior Member, IEEE}%
\thanks{Department Intelligent Transport Systems, Johannes Kepler University Linz, Altenberger Straße 69, 4040 Linz, Austria.
\texttt{\{novel.certad\_hernandez, enrico.del\_re, joshua.varughese, cristina.olaverri-monreal\}@jku.at}}%
}

\newacronym{ADAS}{ADAS}{Advanced Driver Assistance Systems}
\newacronym{ADS}{ADS}{Automated Driving Systems}
\newacronym{SAE}{SAE}{Society of Automotive Engineers}
\newacronym{TOR}{TOR}{Take Over Request}
\newacronym{DDT}{DDT}{Dynamic Driving Task}
\newacronym{ODD}{ODD}{Operational Design Domain}
\newacronym{NDRT}{NDRT}{Non-Driving Related Tasks}
\newacronym{LDS}{LDS}{Laser Distance Sensor}
\newacronym{LIDAR}{LIDAR}{Light Detection And Ranging}
\newacronym{RMSE}{RMSE}{root mean square error}

\newacronym{VRU}{VRU}{Vulnerable Road User}
\newacronym{V2X}{V2X}{Vehicle-to-Everything}
\newacronym{V2P}{V2P}{Vehicle-to-Pedestrian}
\newacronym{P2V}{P2V}{Pedestrian-to-vehicle}
\newacronym{C-V2X}{C-V2X}{Cellular V2X}
\newacronym{PET}{PET}{Post Encroachment Time}
\newacronym{AW}{AW}{acoustic warning}
\newacronym{VW}{VW}{V2P warning}
\newacronym{NW}{NW}{no warning}
\newacronym{ROI}{ROI}{Region of Interest}
\newacronym{NTP}{NTP}{Network Time Protocol}
\newacronym{LORA}{LoRa}{Long Range}
\newacronym{SCR}{SCR}{Speed Change Ratio}
\begin{document}


\maketitle

\begin{abstract}
This study assesses a Vehicle-to-Pedestrian (V2P) collision warning system compared to conventional vehicle-issued auditory alerts in a real-world scenario simulating a vehicle on a fixed track, characterized by limited maneuverability and the need for timely pedestrian response. The results from analyzing speed variations show that V2P warnings are particularly effective for pedestrians distracted by phone use (gaming or listening to music), highlighting the limitations of auditory alerts in noisy environments. The findings suggest that V2P technology offers a promising approach to improving pedestrian safety in urban areas
\end{abstract}


\section{Introduction}\label{sec:intro}

Road traffic accidents are a significant global concern, with a disproportionate number of fatalities and injuries affecting \glspl{VRU}\cite{ELHAMDANI2020102856}. Among the various factors contributing to these accidents, pedestrian distraction, particularly due to smartphone use, has become a critical issue. Studies have shown that a substantial percentage of pedestrians engage with their smartphones while walking, leading to reduced situational awareness, increased risky behavior, and a higher likelihood of near collisions and accidents \cite{ELHAMDANI2020102856}\cite{VAMSHIKRISHNA2024111}. This challenge is further intensified by the fact that traditional warning methods, such as vehicle horns, may not be effective in diverting the attention of distracted pedestrians \cite{ELHAMDANI2020102856}\cite{VAMSHIKRISHNA2024111}\cite{Ni2020}.

Concerning public transport use, pedestrians face significant risks when sharing the road with buses and trams, especially in areas where infrastructure is not optimized for pedestrian safety. This risk is exacerbated by several factors, including pedestrians hurrying to cross the road to catch a bus or tram, or crossing in unsafe locations \cite{Gankhuyag2023}, and bus stops and tramways that lack proper pedestrian facilities, such as footpaths, pedestrian crossings, and safe waiting areas \cite{TZOURAS2020100205}\cite{YENDRA2024107725}.

To address this, the development of advanced systems that can effectively warn distracted pedestrians of potential dangers is essential. \gls{V2P} and \gls{P2V} communication, which allows direct information exchange between vehicles (cars, buses, trams, etc) and pedestrians, offer a promising solution \cite{Hussein2016}\cite{Pan2023}\cite{Chavhan2023}\cite{Zhang2023}\cite{Gu2022}. However, several challenges and research gaps exist in the implementation and evaluation of both communication systems for pedestrian safety:
\begin{itemize}
    \item \textbf{Uncertainty in pedestrian behavior:} Current systems often fail to account for the unpredictable nature of pedestrian movement, leading to inaccurate collision risk predictions\cite{Pan2023}
     \item \textbf{Smartphone limitations:} Relying on smartphones for \gls{V2P} communication raises concerns about battery consumption, channel load due to numerous devices, and how to avoid disturbing non-crossing pedestrians \cite{ELHAMDANI2020102856}.
     \item \textbf{Need for effective warning mechanisms:} Acoustic warnings are not always sufficient, and the practicality of visual warnings on pedestrian's phones needs to be tested \cite{ELHAMDANI2020102856}\cite{Ni2020}.
     \item \textbf{Lack of realistic evaluation:} Many studies rely on simulations or lab settings, with little real-world field tests, and especially field tests that compare the behavior of a pedestrian being alerted by \gls{V2P} communication and traditional warnings (such as car's horn or tram's bell) in similar conditions \cite{VAMSHIKRISHNA2024111}.
\end{itemize}

This paper addresses these gaps by presenting an experimental study that compares the effectiveness of \gls{V2P} collision warnings with traditional methods in alerting distracted pedestrians. The core research question is:

\begin{itemize}
    \item How do distracted pedestrians respond to a V2P-based collision warning system compared to acoustic warnings from a vehicle operating on a fixed track?
\end{itemize}

\begin{figure}[htbp]
\centerline{\includegraphics[width=0.45\textwidth]{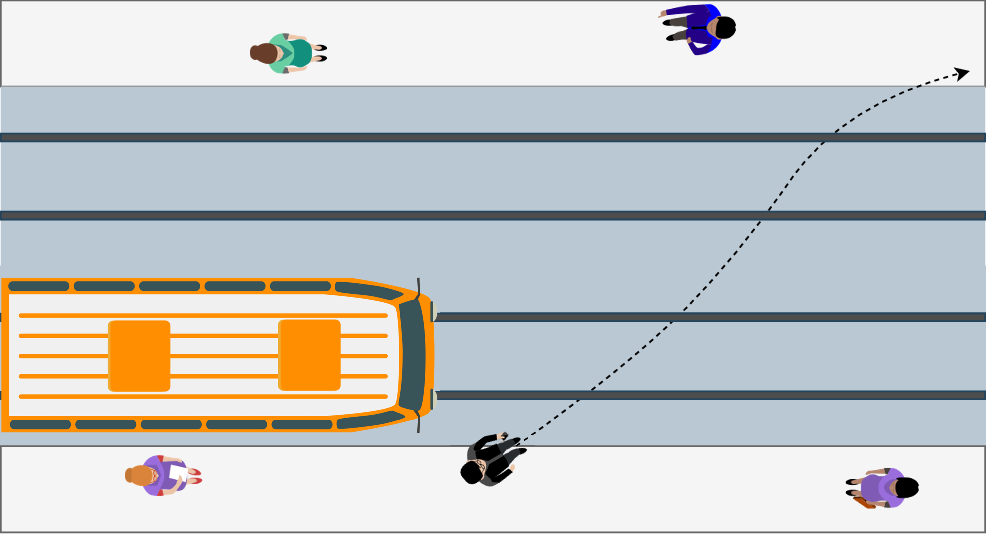}}
\caption{
Scenario replicating urban areas where trams and \glspl{VRU} share the road, and distracted \glspl{VRU} may diagonally cross without noticing an approaching tram.}
\label{fig:scenario}
\end{figure}

To investigate this, we conducted an experiment where participants crossed a replicated shared tram-pedestrian space while using smartphones, simulating common distraction-induced diagonal crossings and reduced awareness of approaching trams (see Fig.~\ref{fig:scenario}). Two conditions were tested to assess pedestrian response. In one condition, a collision warning was issued to the participant’s smartphone via \gls{V2P} communication. In another, an auditory warning was emitted by the vehicle. The participants’ reactions were recorded using a camera mounted on the vehicle, allowing analysis of their visual attention, walking characteristics and overall behavior.

By directly comparing a \gls{V2P} collision warning system and an acoustic warning system, this paper provides empirical evidence for the effectiveness and feasibility of \gls{V2P} technology and its potential to significantly enhance pedestrian safety in urban environments.

The remainder of this paper is organized as follows: the next section describes related studies in the field;  section~\ref{sec:experiment} details the experimental design; section~\ref{sec:evaluation} describes the acquisition and processing of the data collected; sections~\ref{sec:results} presents the results obtained. Finally, section~\ref{sec:conclusions} concludes the present study outlining future work.
\section{Related Work}\label{sec:sota}

Extensive research on \gls{V2P} and \gls{P2V} systems has explored simulation platforms\cite{artal2019}, wireless technologies (\gls{LORA} Wireless network \cite{Pan2023}, Wi-Fi \cite{Chavhan2023}, \gls{C-V2X} \cite{Zhang2023}\cite{Gu2022}) and sensors (LiDAR, cameras) to enable real-time vehicle and pedestrian tracking for effective collision avoidance. However, several studies point out that the effectiveness of these systems can be limited when relying only on the technological aspects\cite{ELHAMDANI2020102856}.

Given the variability of pedestrian behavior, there is a need for \gls{V2P}/\gls{P2V} systems that can adapt to the specific context and behavior of individual pedestrians. Some studies focus on developing algorithms to predict pedestrian trajectories to improve collision risk assessment, but these may not account for all the unpredictable factors in real world scenarios \cite{Chavhan2023}\cite{Hussein2016}\cite{Pan2023}. Some studies show that incorporating perception models in the warning systems and integrating information from multiple sources (vehicle sensors, \gls{V2P} communication, wearable sensors) help to improve reliability 
\cite{olaverri2020}\cite{ELHAMDANI2020102856}.

A significant challenge in pedestrian safety is the growing prevalence of distraction, particularly from smartphone use \cite{HORBERRY2019515}\cite{VAMSHIKRISHNA2024111}. A distracted pedestrian is someone who shifts their focus away from the main task, such as walking or crossing the street, by engaging in another activity \cite{white2017}. Among the technology-based activities reported in the literature as the most common sources of distraction for pedestrians are listening to music, gaming and texting \cite{white2017,JIANG2018170}. Studies have consistently shown that pedestrians engaged with their smartphones exhibit reduced situational awareness, slower walking speeds, increased instances of risky behavior, and a higher likelihood of near collisions\cite{HORBERRY2019515}\cite{VAMSHIKRISHNA2024111}\cite{white2017}
\cite{JIANG2018170}. Additionally, age also plays a role regarding the frequency of being exposed to distraction sources
such as texting or reading on the mobile phone while crossing a road \cite{Schwarz2015}. Naturalistic studies show that distracted pedestrians are less likely to obey traffic signals, check for vehicles, or engage with drivers. Smartphone use while walking leads to weaving, sudden direction changes, and reduced situational awareness, increasing crosswalk violations \cite{HORBERRY2019515}. This suggests that warnings need to be specifically adapted to account for pedestrian distraction \cite{Ni2020}. Therefore, it is important for collision warning systems to be able to adapt to individual pedestrian behavior and various road conditions.

Regarding technological approaches to increasing pedestrian safety, some studies have implemented systems that provide collision warnings. However, they have not extensively analyzed how pedestrian characteristics, such as walking speed, distraction, and visual attention, impact the effectiveness of these warnings compared to more commonly used auditory warnings\cite{Ni2020}\cite{Hussein2016}\cite{morales2019}\cite{demiguel2019}\cite{YUE202050}. The current study contributes to this need by directly comparing the effectiveness of \gls{V2P} warnings with auditory alerts in a real-world setting, while also considering the impact of smartphone distraction on pedestrian behavior.

\begin{figure}[tb]
\centerline{\includegraphics[width=0.48\textwidth]{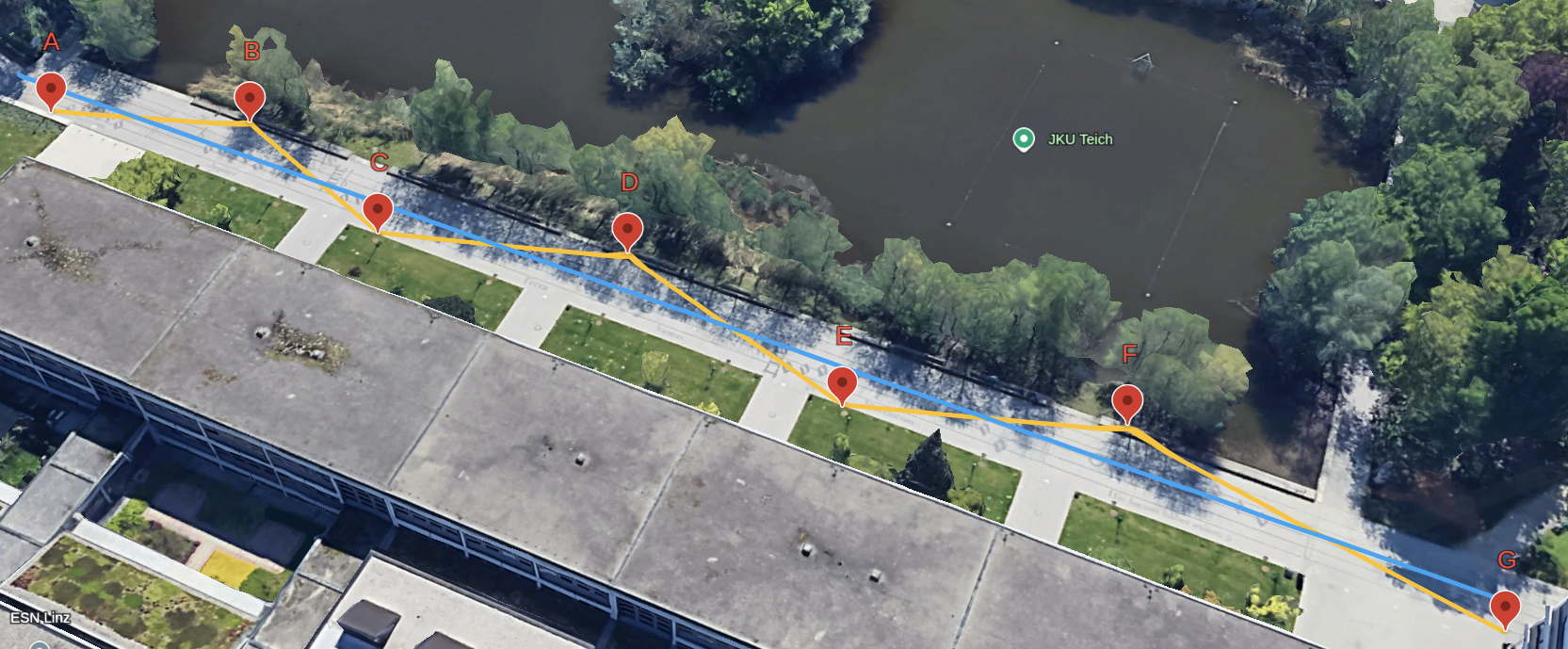}}
\caption{Bird view of the experiment scenario(180x8m)(taken from Google Earth). The red markers indicate the key points, the yellow line represents the intended trajectory for the participants and the blue line exhibits the intended trajectory for the vehicle.}
\label{fig:location}
\end{figure}


\section{Experimental design}\label{sec:experiment}

 An area comprised of a 180-meter-long and 8-meter-wide campus walkway at Johannes Kepler University (Linz, Austria) was selected as the location for the experiment. The area is mostly frequented by pedestrians; however, cyclists and small utility vehicles are also commonly seen. As shown in Fig.~\ref{fig:location}, seven key points were distributed along both sides of the walkway. A sign was physically placed on each key point indicating the corresponding character (from ``A'' to ``G'' ). The participants in the experiment were instructed to walk diagonally through the path connecting the key points from ``A'' to ``G'' and back (yellow line in Fig.~\ref{fig:location}), while the vehicle moved along a fixed track (blue line in Fig.~\ref{fig:location}) in the middle of the walkway. As a result, there were twelve possible crossing points where the pedestrian and the vehicle's paths crossed, simulating the scenario described in Fig.~\ref{fig:scenario}.

Four different agents participating in the experiment were defined:
\begin{itemize}
    \item \textbf{Driver:} This person is in charge of remotely operating the \textbf{vehicle} in a straight line and keeping the right distance ($>$1m) from the \textbf{participant} and explaining the procedure to the \textbf{participant}.
    \item \textbf{Operator:} This person is in charge of selecting and triggering the warnings when the \textbf{participant} enters the defined region of interest. The \textbf{operator} can also send commands (stay and move) to the participant, a detailed explanation can be found in section~\ref{sec:app}.
    \item \textbf{Participant:} Is the subject (pedestrian) whose reaction is under study along the experiment.
    \item \textbf{Vehicle:} The \textbf{vehicle} interacting with the \textbf{participant} along the experiment. To this end, a wheeled robot (Husky A200) was used.
\end{itemize}

A \gls{ROI} was defined in front of the \textbf{vehicle}, measuring 2.6 meters in width (lightly wider than the city trams in Linz) to ensure adequate coverage. The warning was triggered when the participant entered this area.

 As mentioned in section~\ref{sec:sota}, texting, playing games and listening to music have been reported as the most common tasks performed by distracted pedestrians.  In line with this, a trivia application was designed for the experiment and listening to music through headphones was added as a variable condition. In total, two independent variables were controlled during the experiment:
\begin{itemize}
 \item \textbf{Warning Type:} This variable specifies the type of warning issued when the \textbf{participant} entered the vehicle’s \gls{ROI}, resulting in four distinct conditions.
\begin{itemize}
    \item \textbf{\Gls{NW}:} No warning was issued.
    \item \textbf{\Gls{AW}:} An acoustic warning, a bell sound resembling the city's tramway signal in Linz, Austria, was emitted from a vehicle-mounted loudspeaker.
    \item \textbf{\Gls{VW}:} A visual and acoustic warning (visual message and alarm sound) using \gls{V2P} was issued from a server running in the vehicle to the web application in the \textbf{participant}'s smartphone. A detailed description of the implementation is available in section~\ref{sec:app}.
    \item \textbf{Both warning systems (\Gls{AW}+\Gls{VW}):} issued at the same time.
    \end{itemize}
     \item \textbf{Headphones (H):} This variable indicates whether the participant was wearing headphones and listening to music during the experiment:
\end{itemize}

The defined independent variables resulted in eight different combinations which are referenced with the respective acronyms from now on: \gls{NW}, \gls{AW}, \gls{VW}, \gls{AW}+\gls{VW}, \gls{NW}(H), \gls{AW}(H), \gls{VW}(H), and \gls{AW}+\gls{VW}(H).

\subsection{Procedure}\label{sec:procedure}

At the start of the experiment, participants received the following instructions:
\begin{itemize}
    \item Walk at their natural pace ($\sim$1 m/s).
    \item Follow the prompts in the Trivia Application and engage with it continuously while walking or standing.
    \item Proceed to the next key point only when instructed by the Trivia Application.
    \item The vehicle will move along a fixed track without deviation. 
    \item Respond to warnings as they would in real-life situations.
\end{itemize}

\begin{figure}[tbp]
\centerline{\includegraphics[width=0.25\textwidth]{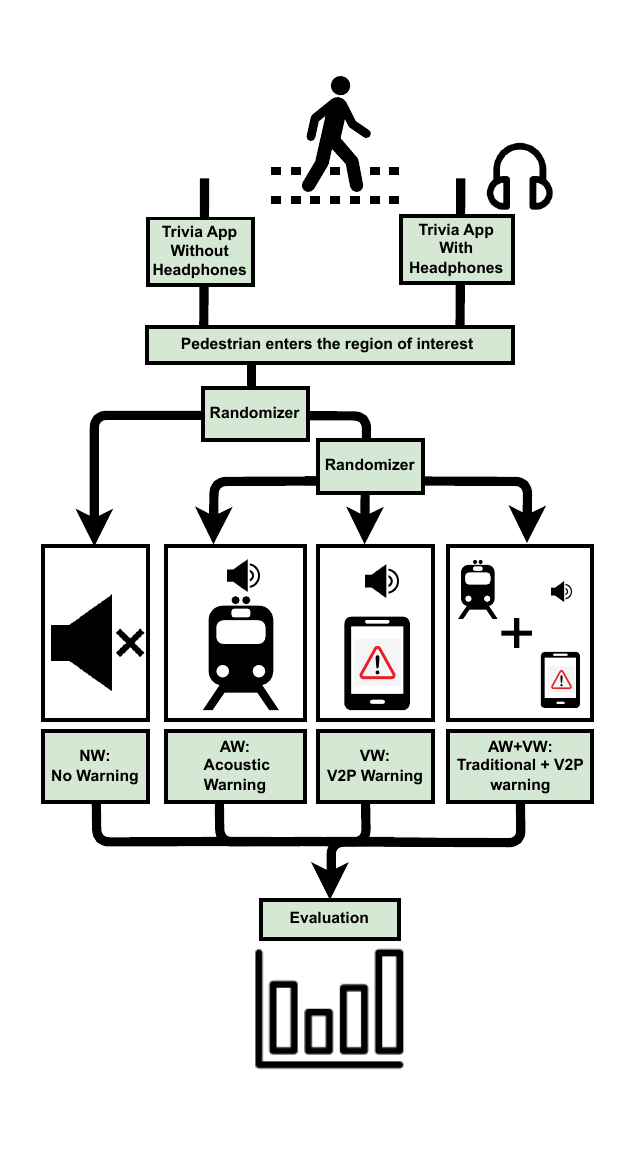}}
\caption{General procedure followed during the experiment.}
\label{fig:procedure}
\end{figure}

The experiment followed the procedure illustrated in Fig.~\ref{fig:procedure}. First, \textbf{participants} were assigned to either the music or no-music condition. They then engaged with the Trivia Application while moving through designated key points. Along the route from point ``A'' to ``G'' and back, there were twelve opportunities to enter the \textbf{vehicle}’s \gls{ROI}. In six instances, no warning was triggered (NW), while in the other six, a warning was issued, with each type (\gls{AW}, \gls{VW} or both) occurring twice. The operator pre-assigned these conditions randomly before the experiment. After completing the first round, \textbf{participants} repeated the experiment under the complementary condition (i.e., with or without music).

The experiment was conducted in three days between December 2024 and January 2025. A total of 10 participants (4 females, 6 males) took part in the recordings. The age distribution was from 20 to 37 years (mean: 29.7, median: 29). 
\subsection{platform}\label{sec:platform}
\begin{figure}[htbp]
\centerline{\includegraphics[width=0.45\textwidth]{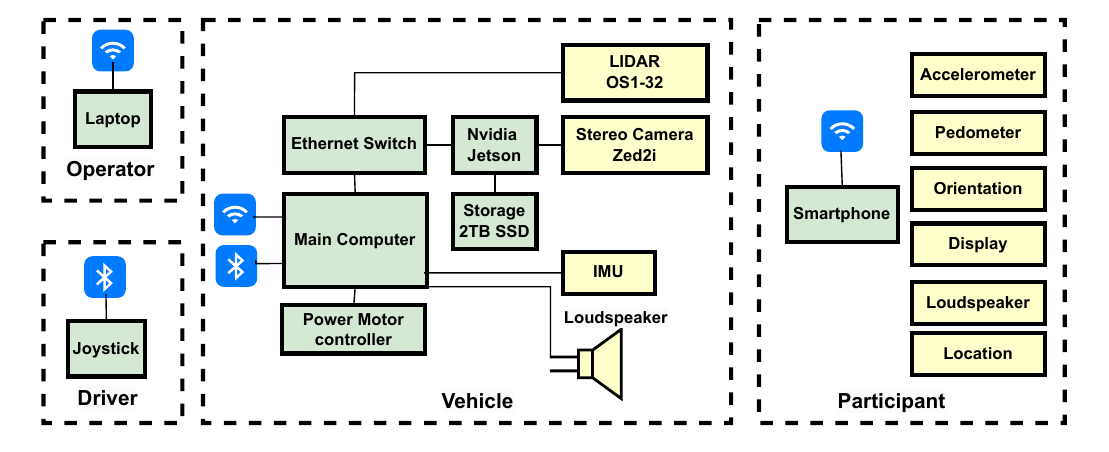}}
\caption{System hardware.}
\label{fig:hardware}
\end{figure}

A general description of the hardware tools utilized for the experiment is shown in the block diagram in Fig.~\ref{fig:hardware}. At the core of the diagram, it is the vehicle for conducting the experiment, a Husky A200. It was equipped with a loudspeaker to play the bell sound intended as a \gls{AW}, and a stereo camera (zed2i) running proprietary object detection algorithms in a Jetson Orion to detect the position and speed of the \textbf{participant} at any time. The vehicle was remotely operated through a Bluetooth joystick. The \textbf{operator} used a laptop connected to the same Wi-Fi network as the \textbf{vehicle}. The \textbf{participant} was given a smartphone, also connected to this network, with its sensors logged via the third-party application Sensor Logger.

The vehicle's odometry sensors and camera readings were acquired using software developed in ROS2. A \gls{NTP} server on the vehicle’s main computer ensured time synchronization with the NVIDIA Jetson, which processed the images. Timestamps were also synchronized with the Web Server (section \ref{sec:app}, as it operated on the vehicle’s main computer. However, smartphone data was synchronized offline by measuring the offset between the \gls{NTP}  server and the smartphone clock before each experimental round.

\subsection{Web Server And Trivia Application}\label{sec:app}
\begin{figure}[ht]
	\centering
        \begin{subfigure}{0.15\textwidth}
		\includegraphics[width=\textwidth]{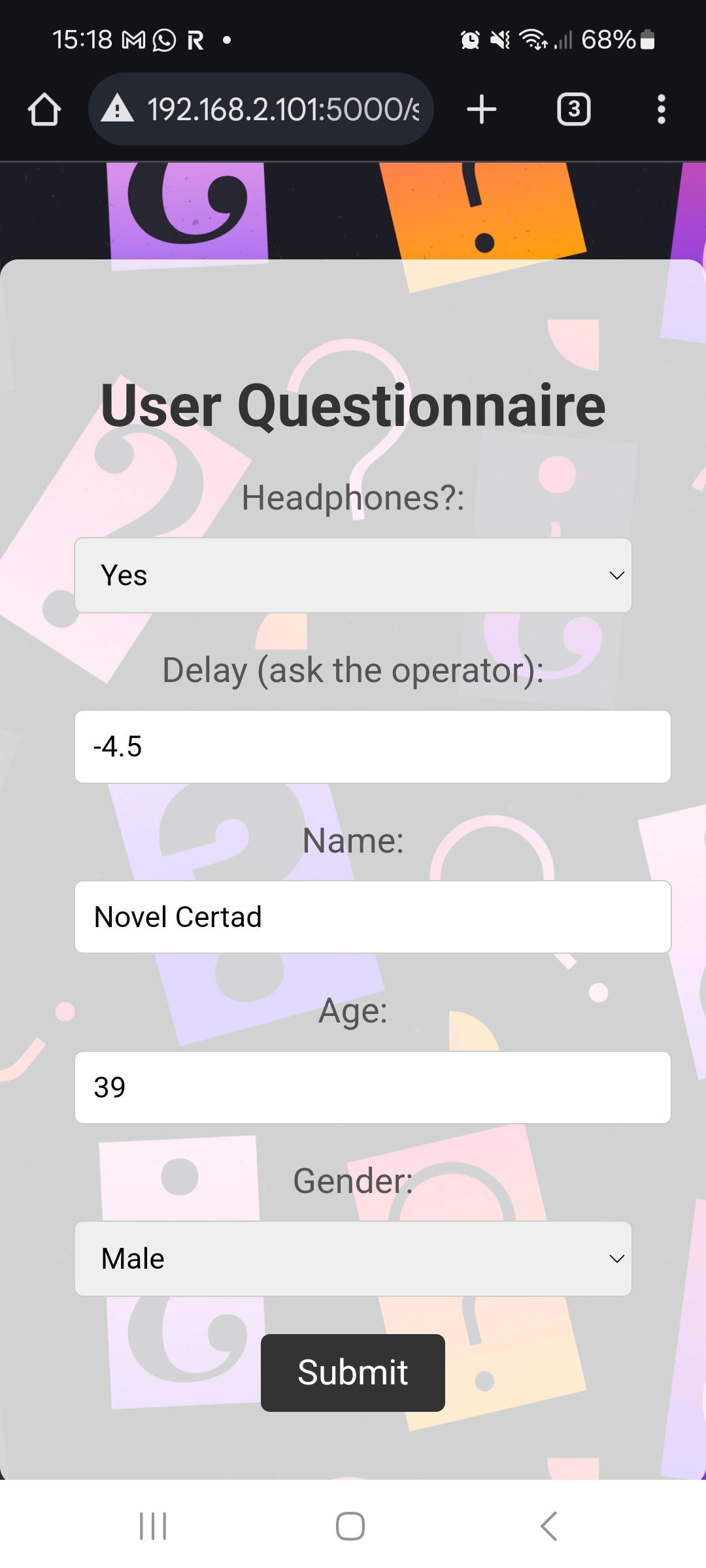}
		\caption{}
	    \label{fig:trivia_app_survey}
	\end{subfigure}
	\begin{subfigure}{0.15\textwidth}
		\includegraphics[width=\textwidth]{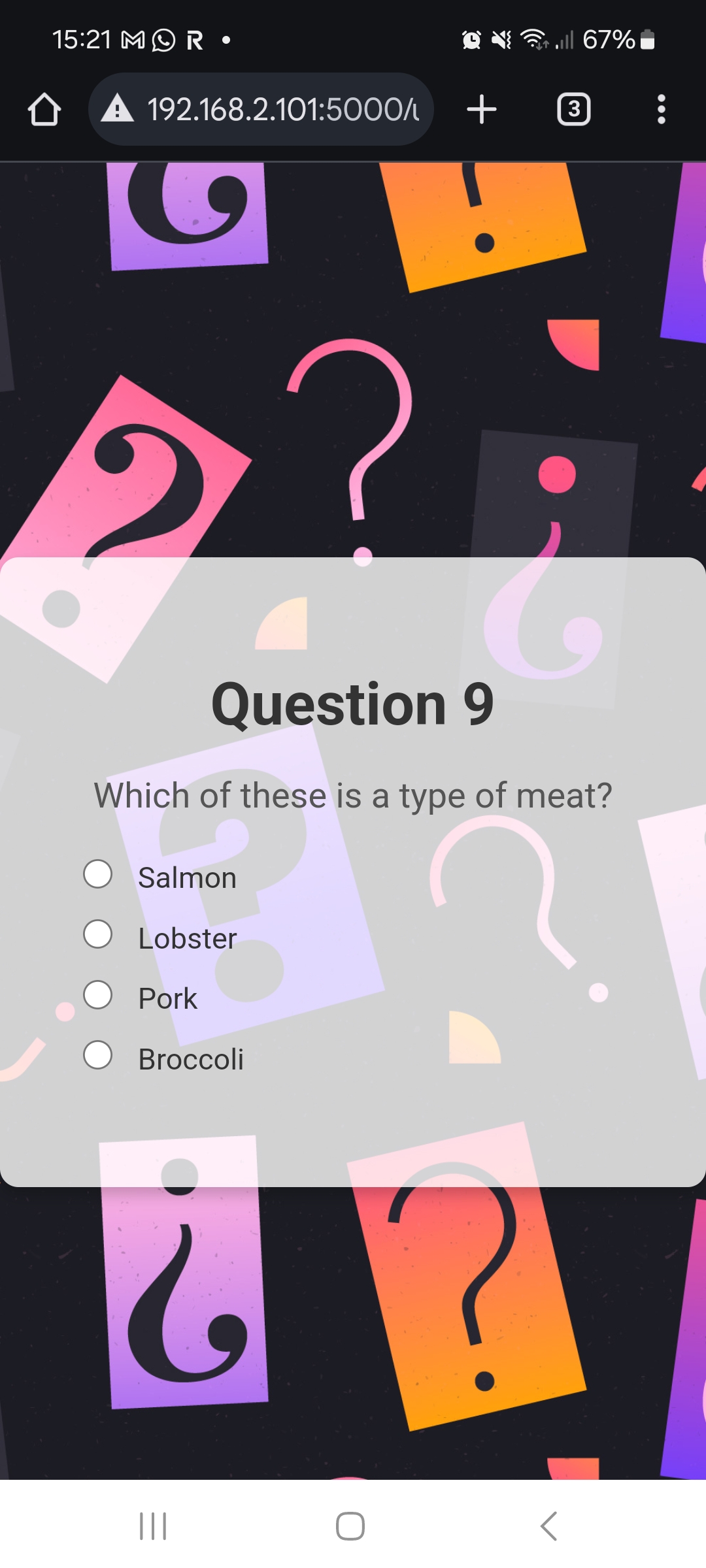}
		\caption{}
	    \label{fig:trivia_app_no_warn}
	\end{subfigure}
	\begin{subfigure}{0.15\textwidth}
		\includegraphics[width=\textwidth]{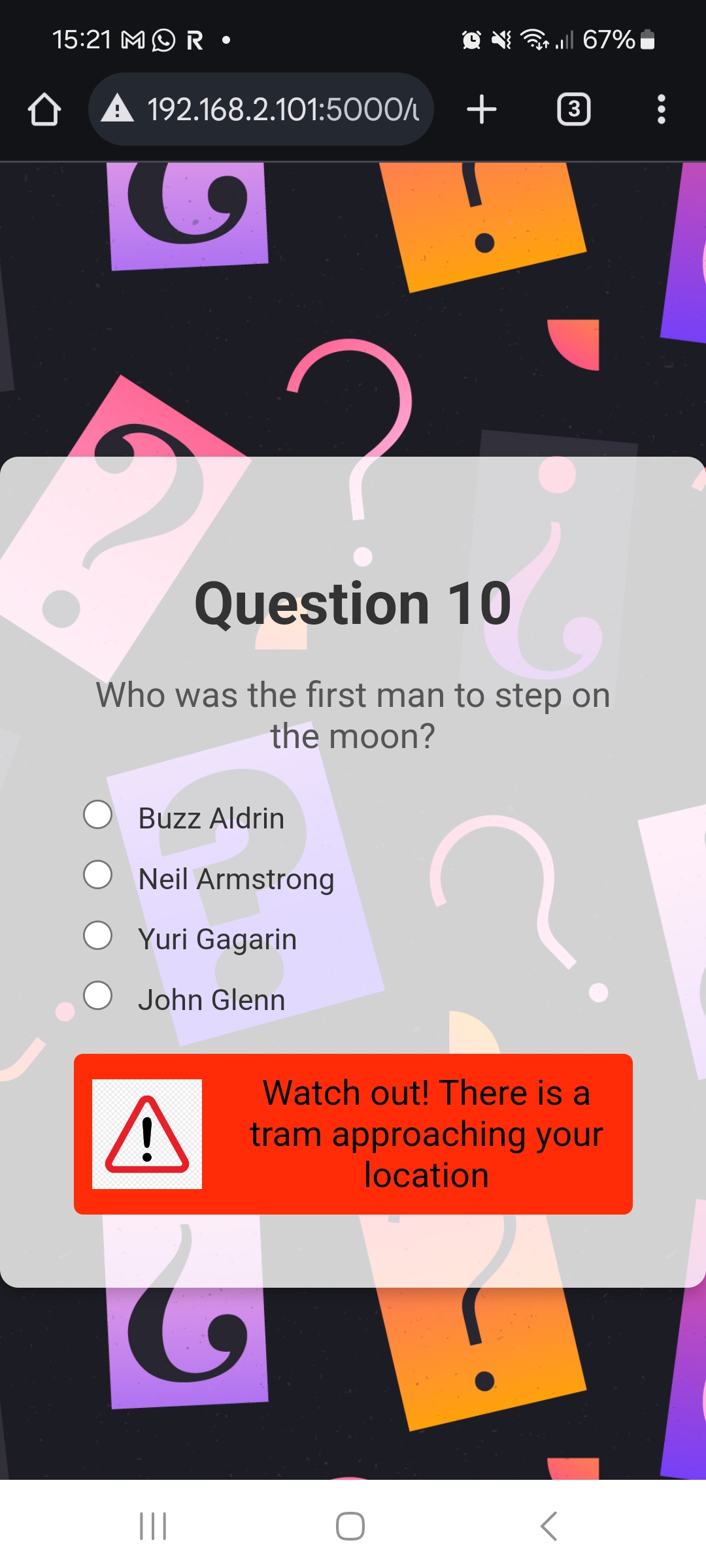}
		\caption{}
	    \label{fig:trivia_app_warn}
	\end{subfigure}
	\caption{Three different screen captures from the trivia application (\textbf{participant} view). (a) Initial survey asking data from the \textbf{participant}. (b) An example of trivia question. (c) An example of trivia question displaying the \gls{V2P} warning (\gls{VW}).}
	\label{fig:trivia_app}
\end{figure}

The application was designed with three key functions: (1) providing an immersive trivia experience to engage \textbf{participants} and divert their attention from the road, (2) serving as the platform for \gls{V2P} warnings via visual alerts (Fig~\ref{fig:trivia_app_warn}) and an alarm sound, and (3) tracking participants' response times.

Upon starting the application, the \textbf{participants} were prompted to complete a short questionnaire that collected their name, age, gender, and whether they were using headphones (Fig.~\ref{fig:trivia_app_survey}). The application's core functionality was a trivia quiz presenting multiple-choice questions (Fig.~\ref{fig:trivia_app_no_warn}) from a database of 500 general knowledge questions (mathematics, geography, culture, literature, history). Questions were displayed one at a time, with real-time response tracking. This data is saved to a CSV file for later analysis, allowing the \textbf{operator} to track the \textbf{participant}'s engagement and performance. 

The \textbf{operator} had additional control through an integrated dashboard, where he could manually issue the warnings (\gls{AW} or \gls{VW}) or the command to instruct the participant to move to next key point.

\begin{figure}[htbp]
\centerline{\includegraphics[width=0.40\textwidth]{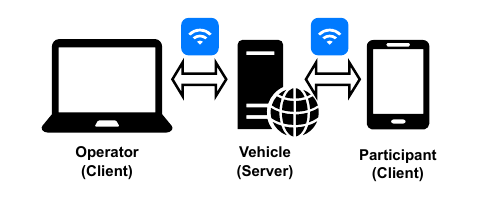}}
\caption{The trivia app runs in a server located in the main computer of the vehicle. Both the \textbf{operator} and the \textbf{participant} connect to the server as clients.}
\label{fig:client-server}
\end{figure}

The trivia application was developed using Flask and Python. As shown in Fig.~\ref{fig:client-server}, the application followed a client-server architecture, with the server running on the \textbf{vehicle}'s main computer. The \textbf{operator} and \textbf{participant}, acting as clients, connected via Wi-Fi and accessed the application through a web browser.

\section{Processing and Evaluation}\label{sec:evaluation}

Pedestrian responses to the different configurations were assessed as follows:

First, camera-detected objects were manually analyzed to exclude non-participant \glspl{VRU}. This process yielded 228 trajectories (12 from one participant and 24 from each of the other nine). The trajectories were then transformed from the vehicle’s local reference frame to a global reference frame.

\begin{figure}[htbp]
\centerline{\includegraphics[width=0.4\textwidth]{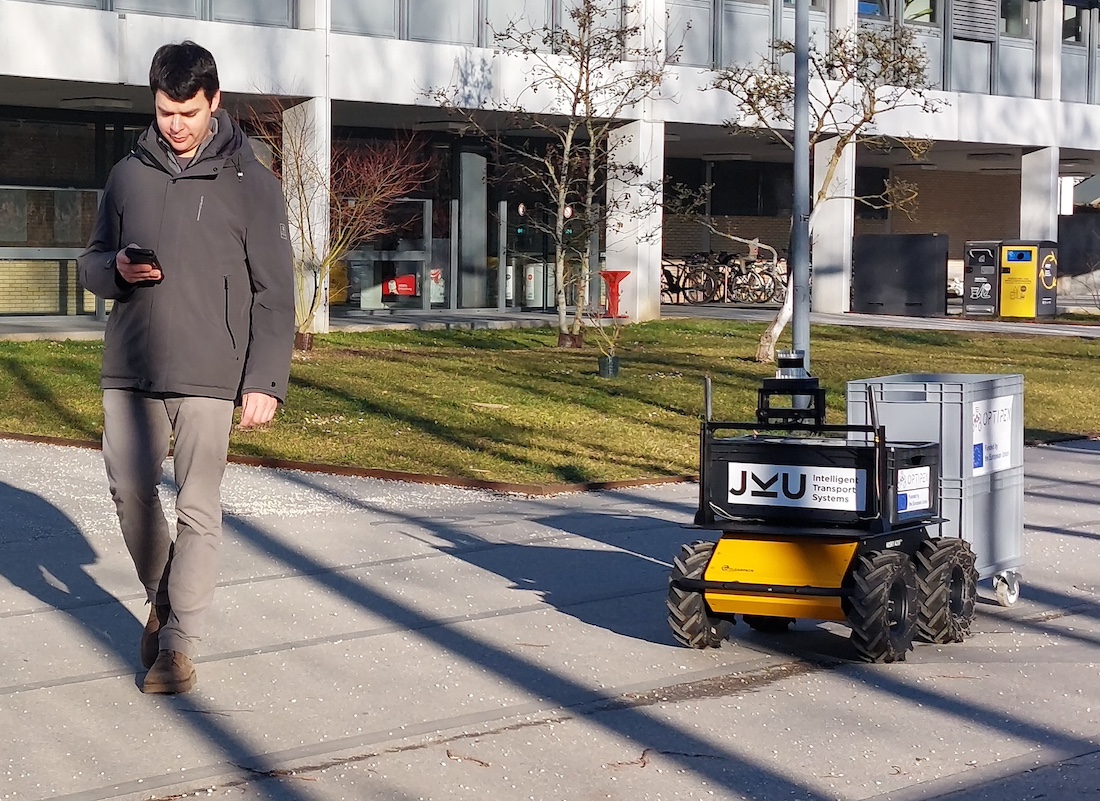}}
\caption{Example of a distracted participant after entering the \gls{ROI}.}
\label{fig:experiment_example}
\end{figure}

To ensure analytical robustness, a filtering process was applied to remove outliers and exclude trajectories with inconsistencies or incomplete data. This included cases where \textbf{participants} were occluded by other pedestrians or the scene contained excessive \gls{VRU}. In addition, crossings where participants stopped to yield to the vehicle were excluded. After filtering, 219 trajectories remained suitable for analysis. An example of the resulting pedestrian crossings is shown in Fig~\ref{fig:experiment_example}

After filtering, key metrics were computed to assess the impact of warning type on participant behavior. The \gls{SCR} was introduced to standardize response comparisons across warning types, accounting for natural variability in pedestrian pace over extended paths. The $SCR_{v}$ was calculated by normalizing the instant speed after entering the \gls{ROI} ($v_{post}(t)$) against the average speed before entering the \gls{ROI} ($\langle v_{pre}(t) \rangle$) as shown in \eqref{eq:scr}

\begin{equation}
SCR_{v}=\frac{v_{post}(t)}{\langle v_{pre}(t) \rangle}
\label{eq:scr}
\end{equation}

Finally, paired t-tests were conducted to assess the statistical significance of differences in $SCR_{v}$ across warning types, verifying whether the observed behavioral changes were significant.

\section{Results}\label{sec:results}
\begin{figure}[htbp]
\centerline{\includegraphics[width=0.45\textwidth]{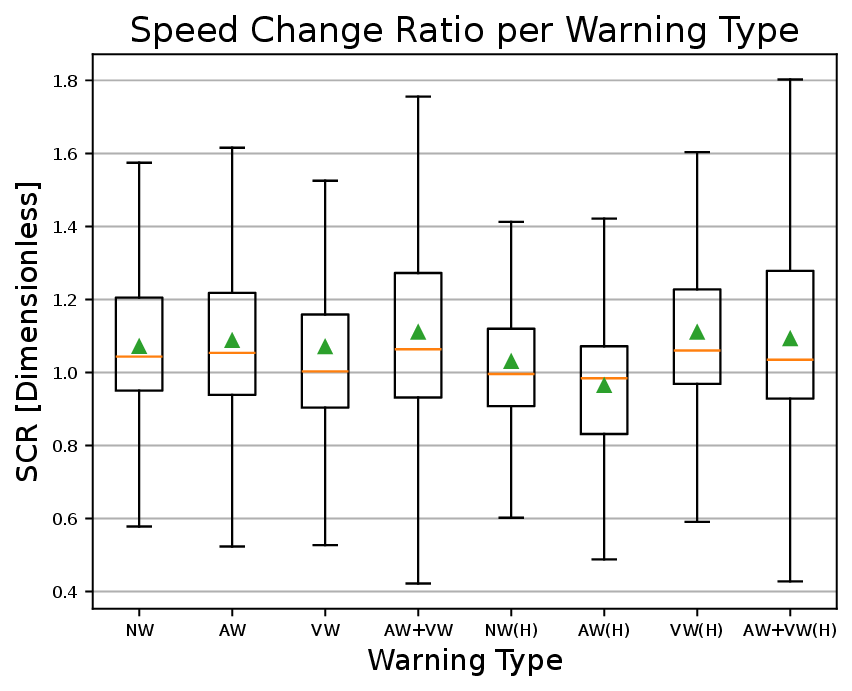}}
\caption{Speed variation before and after pedestrian entry into the Region of Interest (ROI). The boxes extend from the first quartile (Q1) to the third quartile (Q3) of the data. The mean is shown as a green triangle and the median as an orange line.}
\label{fig:speed_variation}
\end{figure}

The numerical values for the mean $\langle SCR_{v} \rangle$ and the standard deviation $SD(SCR_{v})$ are included in the Table~\ref{table:significant}, along with the results from the t-tests comparing all the combinations used in the experiment. The Fig.~\ref{fig:speed_variation} shows a boxplot with the distribution of $SCR_{v}$ per warning type.

\begin{table*}[t]
\caption{Results from applying paired t-student test between all the distributions. P-values$<$0.05 in bold. Last two rows represent the mean ($\langle SCR_{v} \rangle$) and standard deviation (SD $SCR_{v}$) per condition.}
\label{table:significant}
\begin{center}
\begin{tabular}{|c|c|c|c|c|c|c|c|c|}	
\hline
 & \textbf{\gls{NW}} & \textbf{\gls{AW}} & \textbf{\gls{VW}} & \textbf{\gls{AW}+\gls{VW}} & \textbf{\gls{NW}(H)} & \textbf{\gls{AW}(H)} & \textbf{\gls{VW}(H)} & \textbf{\gls{AW}+\gls{VW}(H)} \\
\hline
\gls{NW}            & - & 0.5342 & 0.9654 & 0.1460 & 0.0642 & \textbf{$<$0.0001} & 0.0994 & 0.4465 \\
\hline
\gls{AW}            & 0.5342 & - & 0.5388 & 0.4179 & \textbf{0.0179} & \textbf{$<$0.0001}& 0.3667 & 0.8571 \\ 
\hline
\gls{VW}            & 0.9654 & 0.5388 & - & 0.1648 & 0.1048 & \textbf{$<$0.0001} & 0.1226 & 0.4539 \\ 
\hline
\gls{AW}+\gls{VW}   & 0.1460 & 0.4179 & 0.1648 & - & \textbf{0.0016} & \textbf{$<$0.0001} & 0.9968 & 0.5614 \\ 
\hline
\gls{NW}(H)         & 0.0642 & \textbf{0.0179} & 0.1048 & \textbf{0.0016} & - & \textbf{0.0023} & \textbf{0.0003} & \textbf{0.0185} \\
\hline
\gls{AW}(H)         & \textbf{$<$0.0001} & \textbf{$<$0.0001} & \textbf{$<$0.0001} & \textbf{$<$0.0001} & \textbf{0.0023} & - & \textbf{$<$0.0001} & \textbf{$<$0.0001} \\ 
\hline
\gls{VW}(H)         & 0.0994 & 0.3667 & 0.1226 & 0.9968 & \textbf{0.0003} & \textbf{$<$0.0001} & - & 0.5249 \\ 
\hline
\gls{AW}+\gls{VW}(H)& 0.4465 & 0.8571 & 0.4539 & 0.5614 & \textbf{0.0185} & \textbf{$<$0.0001} & 0.5249 & - \\ 
\hline
\hline
$\langle SCR_{v} \rangle$ & 1.07 & 1.09 & 1.07 & 1.11 & 1.02 & 0.96 & 1.11 & 1.09  \\
\hline
SD $SCR_{v}$    & 0.25 & 0.29 & 0.30 & 0.31 & 0.22 & 0.24 & 0.24 & 0.33 \\
\hline
\end{tabular}
\end{center}
\end{table*}

When comparing the $SCR_{v}$ values when the participants were not wearing headphones (Cases: \gls{NW}, \gls{AW}, \gls{VW} and \gls{AW}+\gls{VW}) no significant differences were found. According to post-experiment comments from the participants, it was highlighted that the noise produced by the vehicle's motor and wheels was at least as louder as the auditory warning (\gls{AW}) itself. In any case the $\langle SCR_{v} \rangle$ resulted in an average increase between 7\% and  11\% in the participant speed after entering the \gls{ROI}.

On the contrary, the $SCR_{v}$ values when the participants were listening to music (Cases: \gls{NW}(H), \gls{AW}(H), \gls{VW}(H) and \gls{AW}+\gls{VW}(H)) presented significant differences that can be analysed in more detail. For \gls{NW}(H), speed increased marginally by 2\% after entering the \gls{ROI}. In contrast, \gls{AW}(H)  showed an average speed decrease of 4\%. This may be due to some participants still perceiving the traditional auditory warning through their headphones. 

Finally, no significant differences were found between \gls{VW}(H) and \gls{AW}+\gls{VW}(H). Participants primarily responded to the \gls{V2P} warning, with an average speed increase of 9–11\%. Across all distributions, $SCR_{v}$  values did not significantly differ between cases where participants were not wearing headphones (\gls{NW}, \gls{AW}, \gls{VW} and \gls{AW}+\gls{VW}) and cases where V2P warnings were issued while participants wore headphones (\gls{VW}(H) and \gls{AW}+\gls{VW}(H)).
\section{Conclusions and Future Work}\label{sec:conclusions}

This study demonstrates the potential of Vehicle-to-Pedestrian (V2P) communication to enhance pedestrian safety, particularly for distracted individuals. 

The findings also highlight the limitations of auditory warnings, which become ineffective when pedestrians wear headphones, emphasizing the need for non-auditory warning systems like \gls{V2P}. Additionally, the observation that pedestrian behavior remained unchanged in the presence of auditory warnings—potentially due to baseline noise from the vehicle—suggests that sound alerts may be less perceptible in real-world conditions. These results underscore the necessity for robust, multimodal safety approaches, especially given the increasing prevalence of distracted walking and the shortcomings of conventional warning methods.

Future research will study the impact on the distance towards the vehicle with a focus on close encounters. It will also evaluate \gls{V2P}  warnings across various distraction scenarios, identifying the most effective alert modality, and assessing strategies to prevent over-reliance while maintaining pedestrian attention to traffic risks.

\section*{Acknowledgment}
The work has been conducted as a part of OptiPEx project (No.101146513) funded by the European Union. Views and opinions expressed are however those of the author(s) only and do not necessarily reflect those of the European Union or CINEA. Neither the European Union nor the granting authority can be held responsible for them.
\bibliographystyle{IEEEtran}
\bibliography{paper}

\end{document}